# Evaluation of pedodiversity and land use diversity in terms of the Shannon Entropy

**Tetsuo Yabuki, Yumi Matsumura, and Yoko Nakatani**
Rakuno Gakuen University, Bunkyodai Midorimachi, Ebetsu, Hokkaido, Japan

E-mail: yabuki@rakuno.ac.jp

**Abstract.** Recently, the Shannon entropy, which was introduced originally as a measure of information amount, has been widely used as a useful index of various diversities such as biodiversity and geodiversity. In this work we have evaluated the diversity of soil and land use, in both composition distributions and spatial distributions, in terms of the Shanonn entropy. Moreover we have also proposed how to estimate the connection between the diversity of soil and land use in the spatial distribution using mutual entropy, and carried out the estimation of its connection.

**Key words:** Shannon entropy, mutual entropy, geodiversity, soil and land use, connection index

**1. Introduction**

In recent times, geodiversity has been of great concern all over the world. Pedodiversity, which is the diversity of soil distribution, is one example of geodiversity and has been an important topic of research in soil science since the 1990s. Several studies[1-7] on this topic have been made. In these studies the Shannon entropy[8,9] was used as an index of the diversity in soil composition distribution. The Shannon entropy which had been introduced originally in the informal theory was applied as a measure of biodiversity by MacArthur[10].Recently Ibáñez et al.[1,2] and Martín and Rey[5]    have suggested the usefulness of the Shannon entropy as a valuable index of diversity. On the other hand, land use has also become an important topic of projects in various fields such as ecological research, land planning, and constructing urban networks[11].

 We have believed that pedodiversity is closely related to biodiversity, and might determine land use at the same time. In this work, we have evaluated the diversities of soil and land use in their composition distributions of various regions, specifically 10 cities of Hokkaido, Japan. And we have proposed a diversity index by which we can evaluate the diversities of soil and land use in their spatial distributions in terms of the Shannon entropy and carried out its evaluation in the same regions. Finally we have also proposed a connection index by which we can evaluate the connection between soil and land use spatial distributions using the mutual

Evaluation of pedodiversity and land use diversity in terms of the Shannon entropy

Entropy[8,12], and carried out its evaluation in the same regions. We have performed our calculation using processed data from a Geographical Information System (GIS) and its database.

## 2. The diversities of soil and land use in their composition distributions and spatial distributions of various regions of Hokkaido, Japan

In this work, we have developed the following formula:

$$Y_A = \frac{-\sum_{k=1}^{n} p_k \log p_k}{\log n} \cdots (1)$$

where n and $p_k$ are defined as follows.

(A) In the case of the diversity of soil and land use in their composition distributions, n represents the number of types of soils or land use, and $p_k$ is defined as the area ratio of kth soil or land use to the total area of each region. In this case, the diversity $Y_A$ indicates diversity in components making up soil or land use composition in each region.

(B) In the case of the diversity of soil and land use in their spatial distribution, n represents the number of spatial meshes, and $p_k$ is defined as the area ratio of kth spatial mesh to the total area of each soil or each land use. In this case, diversity $Y_A$ indicates diversity in spatial distribution of each soil or each land use.

In both cases, the value of diversity index $Y_A$ satisfies the following inequality:

$$0 \leq Y_A \leq 1 \cdots (2)$$

$Y_A$ can take 0 when there is a highly non-uniform distribution of relative abundance, that is, when one or few object dominate, and can take 1 when all species are equiprobable.

### 2.1 *The diversity of soil and land use in their composition distributions*

There are forty six kinds of soils and eleven kinds of land use in Hokkaido, Japan. We calculated composition diversity indices according to the formula (1), in order to evaluate how rich the soil composition of each city in Hokkaido is, and to see how the richness of land use composition of each city has been changed. In Figure 1(a) the changes of land use diversity in its composition distribution for each city from 1976 to 1997 are shown, and in Figure 1(b) the soil diversity in its composition distribution for each city is shown. We have found that land use diversities had gradually changed and increased in almost all cities besides Sapporo, the capital of Hokkaido. We have inferred from the GIS data of Sapporo that because the area of housing land had increased in Sapporo, the land use diversity had decreased in Sapporo. From Figure 1(a) and (b) we can see the roughly positive correlation between the diversities of soil and land use, but we see also there are gaps between them in several cities.

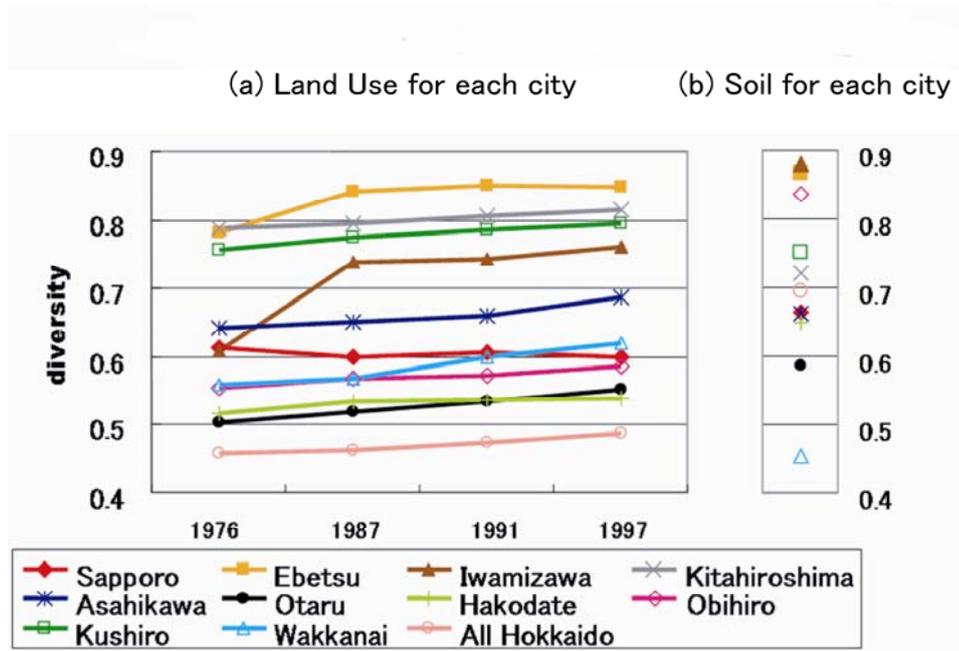

**Figure 1.** The diversity of Land use and soil composition distribution.

Figure 1(a) shows the changes of land use diversity in its composition distribution for each city from 1976 to 1997. Figure 1(b) shows the soil diversity in its composition distribution for each city.

2.2 *The diversity of soil and land use in their spatial distributions*

It can be inferred that the more spatially scattered each soil is, the more spatially scattered land use is, and as a result a network between small elements of land use, such as farms of various crops, can be realized. In order to evaluate how much spatially scattered each soil and land use is, we have calculated spatial diversity indices according to the formula (1). When calculating a spatial diversity index $Y_A$, we set a mesh scale and the value of $Y_A$ depends on its scale.

In Figure 2, an artificial example of spatial distribution of soil is shown. The value of our index $Y_A$ for this artificial distribution indicates 0.87. In Figure 3 and Figure 4, actual examples of spatial distribution of soil in Ebetsu (a city among 10 cities) are shown. In Figure 3,

Evaluation of pedodiversity and land use diversity in terms of the Shannon entropy low spatially scattered soil is shown for which the value of diversity index $Y_A$ indicates 0.867, and on the other hand, in Figure 4, high spatially scattered soil is shown for which the value of diversity index $Y_A$ indicates 0.565.

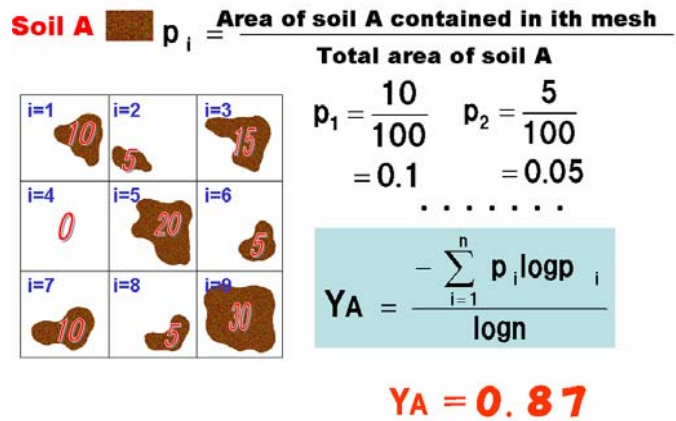

**Figure 2.** An artificial example of spatial distribution of soil.

A spatial distribution of virtual soil A is shown. The diversity of this spatial distribution, that is the degree of spatial scattering, is found as $Y_A=0.870$.

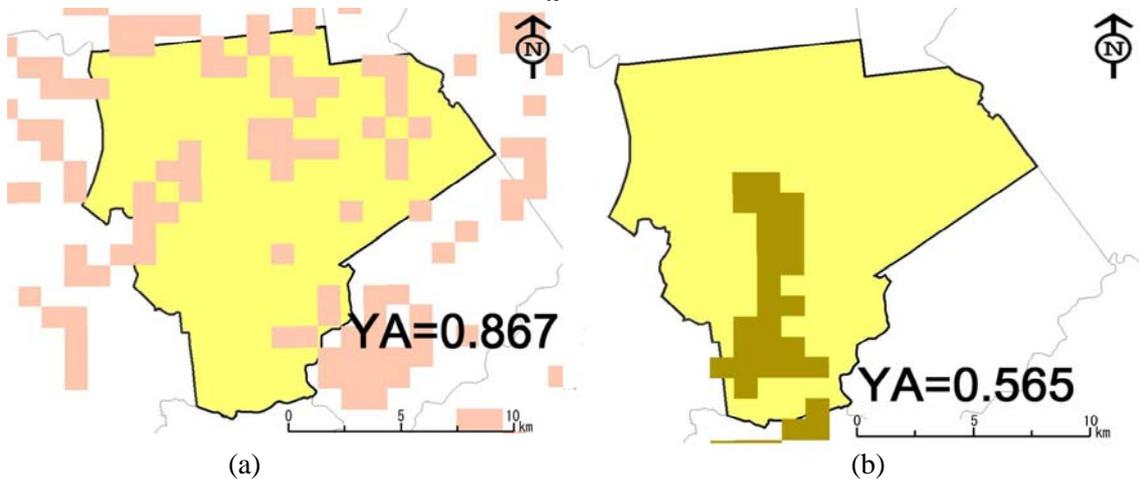

(a)          (b)

**Figure 3.** Actual examples of spatial distribution of soil.

Two examples of spatial distributions of an actual soil in Ebetsu (a city among 10 cities) is shown. Figure 3(a) is an example of higher spatial scattering and Figure 3(b) is an example of lower spatial scattering. The diversities of spatial distributions, namely the degree of spatial scattering, of Figure 3(a) and Figure 3(b) are found as $Y_A=0.867$ and $Y_A=0.565$ respectively.

Evaluation of pedodiversity and land use diversity in terms of the Shannon entropy

We have calculated the average value of spatial diversity index $Y_A$ for all kinds of soils in each city, and we show the result in Figure 4, in which five mesh scales, 5km, 4km, 3km, 2km, and 1km are chosen and shown on the horizontal axis. By comparing Figure 4 with Figure 1(b), we can see that the regional difference of diversity in spatial distribution is less than that of diversity in composition distribution, and we can also see that one city (Ebetsu city) has the highest soil diversity in both composition distribution and spatial distribution.

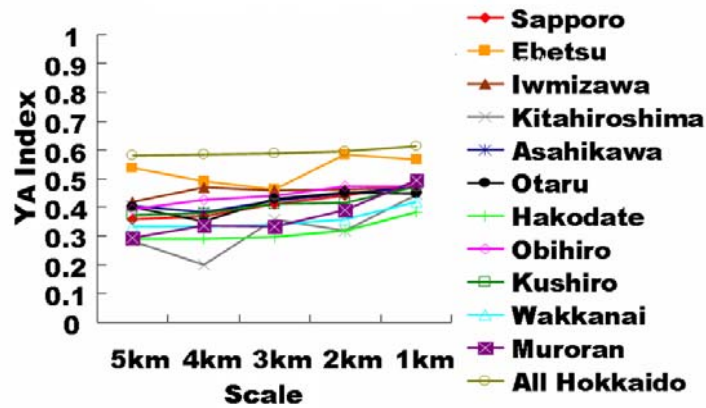

**Figure 4.** The result of the average values of spatial diversity index $Y_A$ in each city.

Average values of spatial diversity index $Y_A$ for all kinds of soils in each city are shown in this Figure, in which four mesh scales, 5km, 4km, 3km, 2km, and 1km are chosen and shown on the horizontal axis.

3. **The connection index between soil and land use spatial distributions in various regions (10 cities) of Hokkaido, Japan**

It can be inferred that the higher a value of pedodiversity in the spatial distribution is, the higher a value of the diversity of land use in the spatial distribution becomes, and a network between small elements of land use, such as farms of various crops, can be realized. This network is expected to make some significant circulation in various industries.

In order to ascertain the above conjecture in which the correlation between diversity of soil and land use in their spatial distribution is supposed, we also proposed a connection index between soil and land use using mutual entropy based on the information theory, and we have evaluated its correlation quantitatively in each region.

The formula is as follows:

$$I(A,B) = S(A) + S(B) - S(A,B) \quad \cdots (3)$$

$$r(A,B) = \frac{1}{2}I(A,B)\left(\frac{1}{S(A)} + \frac{1}{S(B)}\right). \quad \cdots (4)$$

where $A$ and $B$ represent the soil and land use respectively, and $S(A)$ and $S(B)$ represent the entropy of the distribution of soil and land use respectively. $S(A)$, $S(B)$, and $S(A,B)$ are given by the following formulae:

$$S(A) = -\sum_{i=1}^{n} p(A_i) \log p(A_i), \quad \cdots (5)$$

$$S(B) = -\sum_{j=1}^{m} p(B_j) \log p(B_j), \quad \cdots (6)$$

$$S(A,B) = -\sum_{i=1}^{n}\sum_{j=1}^{m} p(A_i, B_j) \log p(A_i, B_j), \cdots (7)$$

where, $p(A_i)$ and $p(B_j)$ are the area ratios of ith soil $A_i$ and jth land use $B_j$ of each city respectively, and $p(A_i, B_j)$, which can be called joint probability, is the area ratio in which soil $A_i$ and jth land use $B_j$ are contained simultaneously.

$S(A,B)$ can be called the entropy of the simultaneous distribution of soil and land use; $I(A,B)$ is the mutual entropy between soil and land use; and $r(A,B)$ represents the connection index between soil and land use.
The connection index $r(A,B)$ satisfies the following inequality:

$$0 \leq r(A,B) \leq 1. \cdots (8)$$

$r(A,B)$ can take 0 when $A$ and $B$ are independent from each other, namely when $p(A_i, B_j) = p(A_i)p(B_j)$ is always satisfied, and can take 1 when $A$ and $B$ are completely connected with each other, namely when $S(A,B) = S(A) = S(B)$ is satisfied.
We have calculated $r(A,B)$ in various regions of Hokkaido, Japan. We have shown our

Evaluation of pedodiversity and land use diversity in terms of the Shannon entropy calculation results in Figure 5. We have seen that the minimum value of $r(A,B)$ is 0.134 (Kitahiroshima city) and the maximum value of $r(A,B)$ is 0.332 (Obihiro city). It is will be complementarily considered in the next section that $r(A,B)$ have generally exhibited rather small values.
.

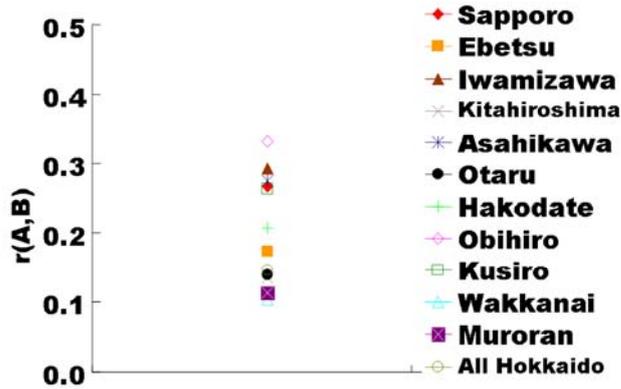

**Figure 5.** The results of connection index between soil and land use distribution for each city.

From these results it can be seen that the connection indexes $r(A,B)$ fall between 0.134 and 0.332. These values are not high. This point will be considered in the discussion of section 3.

### 3. Discussion

3.1 *The diversity in spatial distributions*

In this work, we have originally proposed how to evaluate the degree of soil and land use scattering, that is, the diversity in a spatial distribution. This method has two points to note.

The first point is the mesh scale dependence of diversity in a spatial distribution. An example is shown in Figure 6 and Figure 7, in which two different mesh scales are used. Diversities in two spatial distributions are calculated by the diversity index $Y_A$ in different mesh scales, one of which is given as 1km in Figure 6 and another of which is given as 2km in Figure 7. In the case of the 1km mesh scale (Figure 6) two different distributions have the same diversity index value, while in the case of the 2km mesh scale (Figure 7) two different distributions have different diversity index values, $Y_A = 0$ (minimum value) and $Y_A = 1$ (maximum value), respectively. This feature of index $Y_A$ tells us how large a scale soil or land use is spatially concentrated in.

The second point is the location dependence of diversity in a spatial distribution. An example is shown in Figure 8, in which the same distributions with different locations have entirely different values of diversities. In the case of location given by (a) and (b) of Figure 8, we have the index $Y_A=0$ (minimum value) and $Y_A=1$ (maximum value) respectively. We think this feature is a weak point of index $Y_A$ when it is used as a measure evaluating how much spatial scattering occurred. However, we can address this problem by sliding every mesh in parallel and calculating the average value of index $Y_A$ in all mesh arrangements.

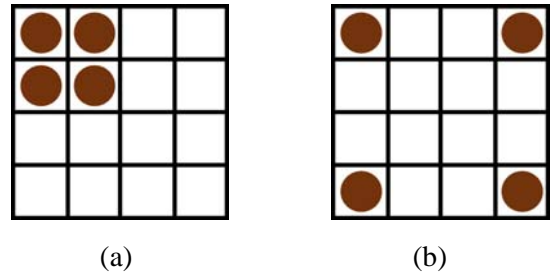

(a)    (b)

**Figure 6.** The mesh scale dependence of diversity in a spatial distribution (1km scale).

In this figure the 1km mesh is chosen. In this mesh scale, a concentrated distribution (a) and a scattered distribution (b) have the same index value, $Y_A=0.5$. We see that in this scale the two distributions (a) and (b) can not be distinguished as the diversity in a spatial distribution.

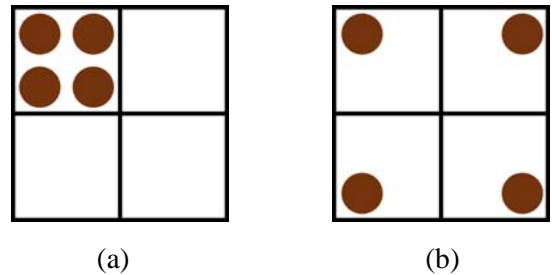

(a)    (b)

**Figure 7.** The mesh scale dependence of diversity in a spatial distribution (2km scale).

In this figure the 2km mesh is chosen. In this mesh scale, a concentrated distribution (a) and a scattered distribution (b) have the different index values, $Y_A=0$ of (a) and $Y_A=1$ of (b). We see that in this scale the two distributions (a) and (b) can be perfectly distinguished as the diversity in a spatial distribution.

Evaluation of pedodiversity and land use diversity in terms of the Shannon entropy

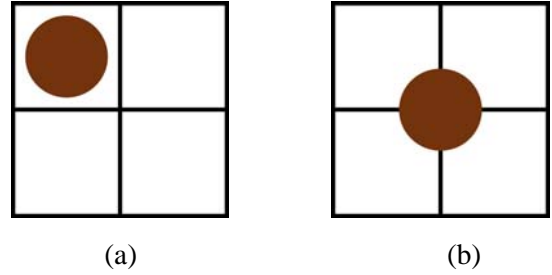

(a)　　　　　　　　(b)

**Figure 8.** The location dependence of diversity in a spatial distribution.

Two figures are shown that even in the same mesh scale and the same spatial distribution, the index $Y_A$ of diversity in a spatial distribution indicates different values, $Y_A=0$ of (a) and $Y_A=1$ of (b).

3.2 *The connection index*

In this work, we have originally proposed how to evaluate the degree of correlation between soil and land use, and we carried out its evaluation using an index which we have called a connection index. After calculating the index in several cases, we have found it generally shows a rather low value(Figure 5) .Hence we examined how large values this index can exhibit in a simple model, in which we have three kinds of soil, $A_1, A_2, A_3$; and three kinds of land use, $B_1, B_2, B_3$; and we supposed the strong connection between $A_1$ and $B_1$, $A_2$ and $B_2$, and $A_3$ and $B_3$. We assumed the following notations and conditions in the simple model of Figure 9.

$$p(A_i, B_i) = a \quad (i = 1,2,3) \quad \cdots (9)$$

$$p(A_i, B_j) = b \quad (i \neq j) \quad \cdots (10)$$

After a little calculation, the connection index $r(A,B)$ in this simple model can be expressed by the following formula:

$$r(A,B) = 1 - \frac{1}{\log 3}\left\{\log(1+2x) - \frac{2x}{1+2x}\log x\right\} \quad \cdots (11)$$

where $x = \dfrac{b}{a}$.

We can see from the behavior of this function (11) shown in Figure 10 that the connection

Evaluation of pedodiversity and land use diversity in terms of the Shannon entropy index $r(A,B)$ rapidly decreases, as x increases only slightly from 0. In the case of Figure 9, we have x=0.2 and $r(A,B)=0.275$. From this point we have seen the connection index $r(A,B)$ exhibits the numerical value near 1 only when the connection between A and B is very strong. Hence we can see that we can evaluate not absolutely but relatively the connection or correlation between A and B using the connection index $r(A,B)$.

We believe the index $r(A,B)$ will be useful in evaluating suitable crops for lands in the future.

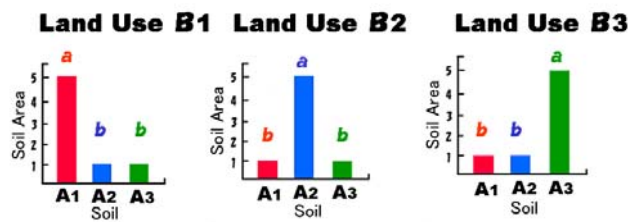

**Figure 9.** A simple model for the connection index $r(A,B)$.

This figure shows a simple model of the connection between land use and soil, in which the joint probabilities are given by $p(A_i, B_i) = a = \dfrac{5}{21}$, $p(A_i, B_j) = b = \dfrac{1}{21}$ $(i \neq j)$.

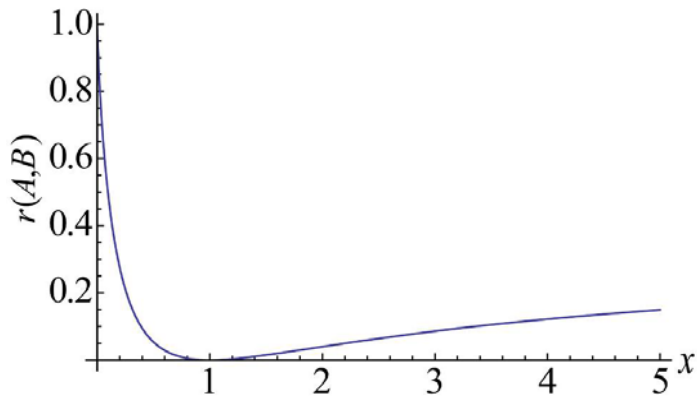

**Figure 10.** The behavior of the connection index $r(A,B)$ in the simple model.

This figure shows the behavior of the connection index $r(A,B)$ in the simple model, where $x = \dfrac{b}{a}$ ($0 \leq x \leq 5$) is shown on the horizontal axis. We can see form this figure that $r(A,B)$ rapidly decreases, as x increases only slightly from 0.

Evaluation of pedodiversity and land use diversity in terms of the Shannon entropy